\documentstyle[12pt]{article}

\def\hybrid{\topmargin -20pt    \oddsidemargin 0pt
        \headheight 0pt \headsep 0pt
        \textwidth 6.25in       
        \textheight 9.5in       
        \marginparwidth .875in
        \parskip 5pt plus 1pt   \jot = 1.5ex}

\hybrid

\catcode`\@=11

\def\marginnote#1{}
\newcount\hour
\newcount\minute
\newtoks\amorpm
\hour=\time\divide\hour by60
\minute=\time{\multiply\hour by60 \global\advance\minute by-\hour}
\edef\standardtime{{\ifnum\hour<12 \global\amorpm={am}%
        \else\global\amorpm={pm}\advance\hour by-12 \fi
        \ifnum\hour=0 \hour=12 \fi
        \number\hour:\ifnum\minute<10 0\fi\number\minute\the\amorpm}}
\edef\militarytime{\number\hour:\ifnum\minute<10 0\fi\number\minute}

\def\draftlabel#1{{\@bsphack\if@filesw {\let\thepage\relax
   \xdef\@gtempa{\write\@auxout{\string
      \newlabel{#1}{{\@currentlabel}{\thepage}}}}}\@gtempa
   \if@nobreak \ifvmode\nobreak\fi\fi\fi\@esphack}
        \gdef\@eqnlabel{#1}}
\def\@eqnlabel{}
\def\@vacuum{}
\def\draftmarginnote#1{\marginpar{\raggedright\scriptsize\tt#1}}

\def\draft{\oddsidemargin -.5truein
        \def\@oddfoot{\sl 2nd draft \hfil
        \rm\thepage\hfil\sl\today\quad\militarytime}
        \let\@evenfoot\@oddfoot \overfullrule 3pt
        \let\label=\draftlabel
        \let\marginnote=\draftmarginnote
   \def\@eqnnum{(\theequation)\rlap{\kern\marginparsep\tt\@eqnlabel}%
\global\let\@eqnlabel\@vacuum}  }

\def\preprint{\twocolumn\sloppy\flushbottom\parindent 2em
        \leftmargini 2em\leftmarginv .5em\leftmarginvi .5em
        \oddsidemargin -.5in    \evensidemargin -.5in
        \columnsep .4in \footheight 0pt
        \textwidth 10.in        \topmargin  -.4in
        \headheight 12pt \topskip .4in
        \textheight 6.9in \footskip 0pt
        \def\@oddhead{\thepage\hfil\addtocounter{page}{1}\thepage}
        \let\@evenhead\@oddhead \def\@oddfoot{} \def\@evenfoot{} }

\def\numberbysection{\@addtoreset{equation}{section}
        \def\theequation{\thesection.\arabic{equation}}}

\def\underline#1{\relax\ifmmode\@@underline#1\else
        $\@@underline{\hbox{#1}}$\relax\fi}

\newskip\humongous \humongous=0pt plus 1000pt minus 1000pt

\newif\ifdtup

\relax

\renewcommand{\theequation}{\arabic{section}.\arabic{equation}}

\newcommand{\be}{\begin{equation}}
\newcommand{\ee}{\end{equation}}
\newcommand{\bea}{\begin{eqnarray}}
\newcommand{\eea}{\end{eqnarray}}

\def\tr{{\mathrm Tr\,}}                  %
\def\Ad{{\mathrm Ad\,}}                  %
\def\ad{{\mathrm ad }}                   %
\begin{document}


\thispagestyle{empty}
\setcounter{page}{0}

\renewcommand{\thefootnote}{\fnsymbol{footnote}}
\fnsymbol{footnote}

${\phantom{\pi}}$
\vspace{.5in}

\begin{center} \Large\bf 
ON THE CHIRAL WZNW PHASE SPACE,\\
EXCHANGE r-MATRICES AND POISSON-LIE GROUPOIDS\footnote{Talk by L. F.
at the S\'eminaire de Math\'ematiques Sup\'erieures, Montr\'eal, 
July 26 - August 6, 1999.}
\end{center}

\vspace{.1in}

\begin{center}
J.~BALOG$^{(a)}$, L.~FEH\'ER$^{(b)}$
and L.~PALLA$^{(c)}$ \\

\vspace{0.2in}

$^{(a)}${\em  Research Institute for Nuclear and Particle Physics,} \\
       {\em Hungarian Academy of Sciences,} \\
       {\em  H-1525 Budapest 114, P.O.B. 49, Hungary}\\

\vspace{0.2in}       

$^{(b)}${\em Institute for Theoretical Physics,
        J\'ozsef Attila University,} \\
       {\em  H-6726 Szeged, Tisza Lajos krt 84-86, Hungary }\\

\vspace{0.2in}   

$^{(c)}${\em  Institute for Theoretical Physics, 
Roland E\"otv\"os University,} \\
{\em  H-1117, Budapest, P\'azm\'any P. s\'et\'any 1 A-\'ep,  Hungary }\\

\end{center}

\vspace{.2in}

{\parindent=25pt
\narrower\smallskip\noindent
ABSTRACT.  This is a review of recent work on the chiral extensions
of the WZNW phase space describing both the extensions 
based on fields with generic monodromy as well as those using 
Bloch waves with diagonal monodromy.  
The symplectic form on the extended phase space is inverted in both cases
and the chiral WZNW fields are found to 
satisfy quadratic Poisson bracket relations 
characterized by monodromy dependent exchange r-matrices.
Explicit expressions for the exchange r-matrices in terms of the
arbitrary monodromy dependent 2-form appearing in the chiral WZNW
symplectic form are given. The exchange r-matrices in the general 
case are shown to satisfy a new dynamical generalization 
of the classical modified Yang-Baxter (YB) equation
and Poisson-Lie (PL) groupoids are constructed that encode this
equation analogously  as PL groups encode the classical YB equation. 
For an arbitrary simple Lie group $G$, exchange r-matrices are 
exhibited that are in one-to-one correspondence with the
possible PL structures on $G$ and admit them as PL symmetries.
}

\newpage

\section{Introduction}
\setcounter{equation}{0}

The Wess-Zumino-Novikov-Witten (WZNW) model \cite{Witt}
of conformal field theory  has proved to be the source 
of interesting structures that play an increasingly important r\^ole
in theoretical physics and in mathematics \cite{DiF,EFV}.
One of the fascinating aspects of the model is that in addition
to its built-in affine Kac-Moody symmetry
it also exhibits certain quantum group properties \cite{qgroup}.
The quantum group properties were originally discovered  in
the quantized model, which raised the question
to find their classical analogues.
The studies at the beginning of the nineties 
led to the consensus that the origin of these quantum group
properties lies in the Poisson-Lie  symmetries
of the so called chiral WZNW phase space that emerges
after splitting the left- and right-moving degrees of 
freedom \cite{Bab}-\cite{AT}.
The  chiral separation arises from the product form of the solution  
of the WZNW field equation given by 
$g(x_L, x_R) = g_L(x_L) g_R^{-1}(x_R)$, 
where $x_C$ $(C=L,R)$ are lightcone coordinates and the $g_C$ 
are quasiperiodic group valued fields with equal monodromies,
$g_C(x+2\pi)=g_C(x) M$ for some $M$ in the WZNW group $G$.
The chiral WZNW Poisson structures found in the literature have the form 
\be 
\Big\{g_C(x)\stackrel{\otimes}{,} g_C(y)\Big\}
={1\over\kappa_C}\Big(g_C(x)\otimes
g_C(y)\Big)\Big(
\hat r + {1\over 2} {\hat I} \,{\mathrm sign}\,(y-x)
\Big), \quad 0< x,y<2\pi,
\label{1}\ee 
where $\hat I$ is given by the quadratic Casimir 
of the simple Lie algebra, ${\cal  G}$,  of the WZNW group, $G$,  and the 
interesting object
is the `exchange r-matrix' $\hat r$.
These classical `exchange algebras' can be regarded as fundamental  
since the current algebra follows as their consequence, and 
help to better understand the quantum group properties 
of the model by means of canonical quantization 
\cite{Chu2,FHT,CL}.
 However, the choice of the chiral Poisson structure is highly non-unique 
due to the fact that the $g_C$ are determined by the physical field $g$ 
only up to the gauge freedom $g_C\mapsto  g_C h$ for any constant $h\in G$.

There are two qualitatively different cases that correspond to  
building the WZNW field out of chiral fields with diagonal 
monodromy (`Bloch waves')
or out of fields with generic monodromy.
For Bloch waves \cite{BDF,Chu5,BBT}, the Poisson structure is essentially 
unique and the associated r-matrix is a solution of the so called classical
dynamical Yang-Baxter (CDYB) equation, which has recently received a lot
of attention \cite{ES}.
For chiral fields with generic monodromy, 
it has been argued in \cite{Gaw,FG} that the possible exchange r-matrices 
should correspond to certain local differential 2-forms $\rho$
on open domains $\check G\subset G$, 
whose exterior derivative is the 3-form that occurs in the WZNW action. 
Until recently, the precise connection between $\rho$ 
and $\hat r$ has not been elaborated, 
and in most papers dealing with generic monodromy actually 
only those very special cases were considered for which $\hat r$ is a 
monodromy independent constant.
 
We here review the main results obtained in our 
recent papers \cite{hosszu,BFP2}, 
where a detailed analysis of the chiral extensions of 
the WZNW phase space was undertaken.
The next section contains an outline of the background to 
the problem.
The subsequent two sections describe the chiral WZNW hamiltonian 
structures in detail.
The final section is devoted to an interpretation of these 
structures in terms of Poisson-Lie groupoids.
  
\section{Chiral extensions of the WZNW phase space}
\setcounter{equation}{0}

We below describe the WZNW Hamiltonian system and the chiral extension 
of its solution space following the spirit of  \cite{Gaw,FG}. 
 
We consider a  simple, real or complex, Lie algebra, ${\cal  G}$,  with a 
corresponding 
connected Lie group, $G$, and identify 
the phase space of the WZNW model associated with the group $G$ as 
\be
{\cal  M}= T^* {\widetilde G}= \{\, (g,J_L)\,\vert\,
g\in \widetilde{G},\,\,\, J_L\in \widetilde{{\cal  G}}\,\},
\label{WZphase1}\ee
where $\widetilde{G}=C^\infty(S^1,G)$ is the loop group and
$\widetilde{{\cal  G}}=C^\infty(S^1,{\cal  G})$ is its Lie algebra.
The isomorphism of the  cotangent bundle $T^* {\widetilde G}$ with 
$\widetilde{G}\times \widetilde{{\cal  G}}$ 
is established by means of right-translations on $\widetilde{G}$.
The elements $g\in \widetilde{G}$ (resp.~$J_L\in \widetilde{{\cal  G}}$) are 
modeled as  $2\pi$-periodic
$G$-valued (resp.~${\cal  G}$-valued) functions on the real line ${\bf R}$.
The phase space is equipped with the symplectic form
\be
\Omega^\kappa= d \int_{0}^{2\pi} d\sigma\, 
{\mathrm Tr}\left( J_L d g g^{-1}\right)
+ {\kappa\over 2 }\int_{0}^{2\pi}d\sigma\,
 {\mathrm Tr}\left(d g g^{-1}\right) \wedge \left(d g g^{-1}\right)'
\label{WZsymp1}\ee
with some constant $\kappa$.
Here prime denotes derivative with respect to the space variable, 
$\sigma\in {\bf R}$,
and for any $A, B\in {\cal  G}$ $\tr(AB)$ denotes a fixed multiple of the 
Cartan-Killing form on ${\cal  G}$.
If $T_\alpha$ and $T^\alpha$ ($\alpha = 1,\ldots, {\mathrm dim}\,{\cal G}$) 
are dual bases of ${\cal G}$,
$\tr(T_\alpha T^\beta)=\delta_{\alpha}^\beta$, 
then $\tr (AB)= A_\alpha B^\alpha$
with $A_\alpha = \tr(A T_\alpha)$, $B^\alpha =\tr(B T^\alpha)$ 
and the usual summation convention in force. 
For the wedge product we use the conventions in \cite{AM}.

Although the expression of $\Omega^\kappa$ appears rather formal at first 
sight,
it can be used to unambiguously associate hamiltonian vector fields 
and Poisson brackets (PBs)
with a set of admissible functions, which include, for example, the 
Fourier components of 
the WZNW field $g$, the `left-current'  $J_L$ and  
the `right-current'  $J_R$ given by    
\be
J_R=  - g^{-1} J_L g + \kappa g^{-1} g^\prime.
\label{I1}\ee
The currents are the 
momentum maps that generate two commuting actions of $\widetilde {G}$ 
on ${\cal  M}$ that correspond respectively to left- and right-translations 
on $\widetilde{G}$.
This means that the following local PB relations are valid:
\bea
&&\{\,\tr (T_\alpha J_L)(\sigma) \,, \tr (T_\beta J_L)(\bar \sigma) \,\}_{WZ}
     =\tr ([T_\alpha,T_\beta] J_L)(\sigma) \delta 
                +\kappa\, \tr (T_\alpha  T_\beta) \, \delta' \nonumber\\
&&\{\,\tr (T_\alpha J_R)(\sigma) \,,  
\tr (T_\beta\, J_R)(\bar \sigma) \,\}_{WZ}
     =\tr ([T_\alpha,T_\beta] J_R)(\sigma) \delta
                -\kappa\, \tr (T_\alpha  T_\beta) \, \delta'\nonumber\\
&&\{\,g(\sigma)\,,  \tr(T_\alpha J_L)(\bar\sigma) \,\}_{WZ}
     =  T_\alpha g(\sigma) \,\delta \nonumber\\
&&\{\,g(\sigma)\,,   \tr(T_\alpha J_R)(\bar\sigma) \,\}_{WZ}
     =  -g(\sigma)T_\alpha \,\delta, 
\eea
together with $\{ J_L(\sigma), J_R(\bar \sigma)\}_{WZ}=0$,
where $\delta :=\delta(\sigma -\bar \sigma)=
{1\over 2\pi} \sum_{n\in {\mathbf Z}} e^{in(\sigma -\bar \sigma)}$.
These PBs can be derived from the  
symplectic form $\Omega^\kappa$, whose precise meaning  
is explained in several papers (see e.g. \cite{3REF}).
Thus we need not dwell on this point, but  
note that in the case of a complex 
Lie algebra the admissible functions depend holomorphically 
on the matrix elements of $g$, $J_L$, $J_R$ in the finite dimensional 
irreducible
representations of $G$, and $\widetilde{G}\times 
\widetilde{{\cal  G}}$ is then 
a model of the holomorphic cotangent bundle.  

The phase space ${\cal  M}$ represents the initial data for the 
WZNW system, whose
dynamics is generated by the Hamiltonian
\be
H_{WZ}={1\over 2\kappa} \int_0^{2\pi}d\sigma\, 
\tr\left(J_L^2 + J_R^2 \right).
\ee
Denoting time by $\tau$ and introducing lightcone coordinates as
\be
x_L := \sigma + \tau,
\quad x_R:=\sigma -\tau,
\quad
\partial_{L}= {\partial \over \partial x_L}= {1\over 2} 
( \partial_\sigma + \partial_\tau),
\quad
\partial_R={\partial \over \partial x_R}= {1\over 2} 
( \partial_\sigma - \partial_\tau),
\ee
Hamilton's equation can be written in the alternative forms \cite{Witt}
\be
\kappa\partial_L g = J_L g, \quad \partial_R J_L=0
\qquad{\Leftrightarrow}\qquad
\kappa\partial_R g = gJ_R, \quad \partial_L J_R=0.
\ee

Let ${\cal  M}^{sol}$ be the space of solutions of the WZNW system.
${\cal  M}^{sol}$ consists of the smooth $G$-valued functions
$g(\sigma, \tau)$ which are $2\pi$-periodic in $\sigma$
and satisfy $\partial_R (\partial_L g\, g^{-1})=0$.
The general solution of this  evolution equation can be written as
\be
g(\sigma, \tau)= g_L(x_L) g_R^{-1}(x_R),
\label{gsol}\ee
where $(g_L, g_R)$ is any pair of $G$-valued, smooth,
quasiperiodic function on ${\bf R}$ with {\em equal monodromies}, i.e.,
for $C=L,R$ one has 
$g_C(x_C + 2\pi) = g_C(x_C)M$ 
with some $C$-independent  $M\in G$.
To elaborate this representation of the solutions in more detail,
we define the space $\widehat {\cal  M}$:
\be
\widehat{\cal  M}:= \{ (g_L, g_R) \vert g_{L,R} \in C^\infty({\bf R}, G),
\quad g_{L,R}(x + 2\pi) = g_{L,R}(x) M
\quad  M\in G\}.
\ee
There is a free right-action of $G$ on $\widehat {\cal  M}$ given by
\be
G\ni h: (g_L, g_R) \mapsto (g_L h, g_R h).
\label{bundle}\ee
Notice that $\widehat {\cal  M}$ is a {\em principal fibre bundle} over
${\cal  M}^{sol}$ with respect to the above action of $G$.
The projection of this bundle,
$\vartheta: \widehat{\cal  M} \rightarrow {\cal  M}^{sol}$, is given by
\be
\vartheta: (g_L, g_R)\mapsto g=g_L g_R^{-1}
\qquad\hbox{i.e.}\qquad g(\sigma,\tau)=g_L(x_L) g^{-1}_R(x_R).
\label{theta}\ee

We can identify ${\cal  M}$ with ${\cal  M}^{sol}$ by associating
the elements of the solution space with their initial data at $\tau=0$.
Formally, this  is described by the map 
$\iota: {\cal  M}^{sol} \rightarrow {\cal  M}$,
\be
\iota: {\cal  M}^{sol}\ni  g(\sigma, \tau) \mapsto
\Bigl(g(\sigma, 0), J_L(\sigma)=(\kappa\partial_L g\, g^{-1})(\sigma,0)\Bigr)
\in {\cal  M}.  \ee
Obviously,
$\iota^* (\Omega^\kappa)$ is then the natural symplectic form on 
the solution space.
Explicitly, 
\be
(\iota^* \Omega^\kappa)(g)=
-\kappa\left( d\,\int_{0}^{2\pi}d\sigma\,  
{\mathrm Tr}\left( g^{-1}\partial_R g\, g^{-1}dg \right)
+{1\over 2 }\int_0^{2\pi} d\sigma\,
 {\mathrm Tr}\left(g^{-1}dg \right) \wedge 
\partial_\sigma \left(g^{-1}dg \right)
\right)\big\vert_{\tau=0}
\label{solsymp}\ee
Regarding now ${\cal  M}^{sol}$ as the base of the bundle
$\vartheta: \widehat{{\cal  M}} \rightarrow {\cal  M}^{sol}$, we obtain
a closed 2-form, $\widehat \Omega^\kappa$, on $\widehat{\cal  M}$,
$\widehat \Omega^\kappa := \vartheta^* (\iota^* \Omega^\kappa)$.
By substituting the explicit formula (\ref{theta}) of $\vartheta$,
one finds 
\be
\widehat\Omega^\kappa(g_L, g_R)= \kappa_L \Omega_{chir}(g_L) +
\kappa_R \Omega_{chir}(g_R),
\quad\hbox{with}\quad
\kappa_L := \kappa, \quad
\kappa_R:= -\kappa,
\label{*}\ee
where $\Omega_{chir}$ is the so called chiral WZNW 2-form:
\bea
&&\Omega_{chir}(g_C)=
- {1\over 2 }\int_0^{2\pi}dx_C\,
 {\mathrm Tr}\left(g_C^{-1}dg_C \right) \wedge \left(g_C^{-1}dg_C \right)'
-{1\over 2} \tr \left( (g_C^{-1} dg_C)(0)
\wedge dM_C 
{\scriptstyle\,} M_C^{-1}\right),\nonumber\\
&& M_C=g_C^{-1}(x) g_C(x+2\pi).
\label{**}\eea
This crucial formula of $\widehat\Omega^\kappa$ was first 
obtained by Gawedzki \cite{Gaw}.

It is clear from its definition 
that $d\widehat \Omega^\kappa=0$, but $\widehat \Omega^\kappa$ is not
a symplectic form on $\widehat{{\cal  M}}$, since it is degenerate.
Of course,  its restriction
to any (local) section of the bundle
$\vartheta: \widehat{{\cal  M}} \rightarrow {\cal  M}^{sol}$ is 
a symplectic form,
since such sections yield (local) models of ${\cal  M}^{sol}$.
On the other hand,  
one can check that  $\Omega_{chir}$ has a non-vanishing
exterior derivative \cite{Gaw}: 
\be
d\Omega_{chir}(g_C)= -{1\over 6} 
\tr\left( M_C^{-1}dM_C \wedge M_C^{-1}dM_C  \wedge 
M_C^{-1} dM_C \right).
\label{dOchir}\ee
Although this cancels from $d\widehat \Omega^\kappa$, since $M_L=M_R$ 
for the elements of $\widehat {\cal  M}$,
it makes the chiral separation of the WZNW degrees of freedom a very
non-trivial problem.

The idea of the chiral separation arises from the observation
that the currents $J_C$ {\em almost} completely determine   
the chiral WZNW fields $g_{C}$, and thus also  $g=g_L g_R^{-1}$, 
by means of the differential equations
\be
\kappa_C \partial_C g_C = J_C  g_C \quad\hbox{for}\quad  C=L,R. 
\ee
Thus it appears an interesting possibility to construct the WZNW model 
as a reduction of a simpler model, in which the left- and right-moving
degrees of freedom would be separated in terms of {\em completely 
independent} 
chiral  fields $g_L$ and $g_R$  
regarded as fundamental variables. 
It is clear that the solution space of such a chirally extended model 
must be 
a  direct product of two identical but independent spaces, i.e., it must 
have the form   
\be
\widehat{\cal  M}^{ext} := {\cal  M}_L \times {\cal  M}_R,
\ee
\be
{\cal  M}_{C}:= \{ g_{C} \,\vert\, g_{C} \in C^\infty({\bf R}, G),
\quad g_{C}(x + 2\pi) = g_{C}(x) M_{C}
\quad  M_{C}\in G\}.
\ee
Ideally, one would like  to endow the space $\widehat {\cal  M}^{ext}$ with 
a symplectic structure, $\widehat \Omega^\kappa_{ext}$, 
 that  reduces to 
$\widehat \Omega^\kappa$ on the submanifold 
$\widehat {\cal  M}\subset \widehat{\cal  M}^{ext}$ defined by the periodicity 
constraint $M_L=M_R$.
It is easy to see that these requirements force 
$\widehat \Omega_{ext}^\kappa$
to have the following form:
\be
\widehat \Omega_{ext}^\kappa(g_L, g_R)= 
\kappa_L \Omega_{chir}^{\rho}(g_L) +\kappa_R \Omega_{chir}^{ \rho}(g_R),
\ee
\be
\Omega_{chir}^{\rho}(g_C) = \Omega_{chir}(g_C) + \rho(M_C) 
\label{Orho}\ee
with some 2-form $\rho$ depending {\em only} on the monodromy of $g_C$.
Since in  the extended model the factors  
$({\cal  M}_C, \kappa_C \Omega_{chir}^{\rho})$ should be {\em symplectic} 
manifolds {\em separately}, the condition
\be
d\Omega_{chir}^{ \rho} = -{1\over 6} 
\tr\left( M_C^{-1} dM_C \wedge M_C^{-1}dM_C  \wedge 
M_C^{-1} dM_C \right) + d\rho(M_C) =0
\label{dOrho}\ee
arises. 
But then we have to face the problem that no globally 
defined smooth 2-form exists on $G$ that would satisfy this condition 
for all $M_C\in G$.  
 
There are two rather different way-outs from the above difficulty \cite{FG}.
The first is to restrict the possible domain of the monodromy 
matrix $M_C$ to some open submanifold in $G$ on which an appropriate 
2-form $\rho$ may be found. 
We refer to a choice of such a domain and 2-form  $\rho$ 
as {\em a chiral extension of the WZNW system} with generic monodromy.

The second possibility is to restrict the domain of the allowed monodromy
matrices much more drastically from the beginning, in such a way that after
the restriction $d\Omega_{chir}$ vanishes, whereby the difficulty disappears.
For example, one may 
achieve this by restricting the monodromy matrices to vary
in a fixed maximal torus of $G$, which amounts to constructing 
(a subset of) the 
solutions of the WZNW field equation in terms of chiral `Bloch waves'.
This second possibility is especially natural in the case of compact 
or complex Lie groups, for which there is only one maximal torus
up to conjugation.
Geometrically, 
the restriction to Bloch waves corresponds to taking a (local) section 
of the bundle $\vartheta: \widehat {\cal  M} \rightarrow {\cal  M}^{sol}$.

\section{Hamiltonian structures for generic monodromy} 
\setcounter{equation}{0}

We here investigate the chiral WZNW phase space
${\cal  M}_C$ introduced above. 
The analysis is the same for both chiralities, $C=L,R$, and we 
simplify our notation by putting ${\cal  M}_{chir}$ 
for ${\cal  M}_C$  and $g$, $M$, $\kappa$ for $g_C$, $M_C$, $\kappa_C$, 
respectively. Thus 
  ${\cal  M}_{chir}$ is parametrized by the  $G$-valued, smooth, 
quasiperiodic field $g(x)$ 
satisfying the monodromy condition 
\be 
g(x+2\pi)=g(x)M \qquad M\in G.  
\label{Moncon}\ee 
The corresponding chiral current, 
$J(x)=\kappa g^\prime(x)g^{-1}(x)\in{\cal G}$,
is a smooth, $2\pi$-periodic function of $x$. 
We then consider a 2-form $\rho$ on a domain $\check G \subset G$,
for which we assume that $d\rho(M)= {1\over 6} 
\tr (M^{-1} dM)^{3\wedge}$ and let 
$\check {\cal  M}_{chir}\subset {\cal  M}_{chir}$ be the set 
of chiral WZNW fields whose monodromy matrix lies in $\check G$.
It turns out that $\kappa \Omega_{chir}^\rho$ defines a
symplectic structure on $\check {\cal  M}_{chir}$ 
if a further condition holds for the pair $(\check G, \rho)$.
In order to describe this condition let us introduce the 
parametrization   
\be 
\rho(M)= {1\over2}\,q^{\alpha\beta}(M){\rm Tr}
\big(T_\alpha M^{-1}dM\big)\wedge{\rm Tr}\big(T_\beta M^{-1}dM\big),
\qquad
q^{\alpha\beta}=-q^{\beta\alpha}, 
\label{qpara}\ee  
with $T_\alpha$ denoting a basis of ${\cal  G}$, whose dual basis 
with respect to $\tr$ is $T^\alpha$.
For any $M\in \check G$,
then also introduce the linear operator $q(M): 
{\cal  G}\rightarrow {\cal  G}$ by
\be
q(M): T^\beta \mapsto q^{\alpha\beta}(M) T_\alpha,
\ee
as well as its shifts, $q_\pm (M):= q(M) \pm {1\over 2}I$, by 
the identity operator $I$.
The further condition that we need is that 
\be
\det \left( q_+(M)- q_-(M) \circ \Ad M^{-1} \right)\neq 0
\qquad
\forall M\in \check G.
\label{detcond}\ee
This condition will guarantee the (weak) non-degeneracy 
of $\kappa\Omega^\rho_{chir}$ on $\check {\cal  M}_{chir}$.
 
To use $\kappa \Omega_{chir}^\rho$ in practice we need to 
establish some notation for
tangent vectors $X[g]$ at $g\in {\cal  M}_{chir}$ and vector fields $X$ over
the chiral phase space. To
this end we consider smooth curves on ${\cal  M}_{chir}$ described by 
functions $\gamma(x,t)\in G$  satisfying 
 \be
\gamma(x+2\pi,t)=\gamma(x,t)M(t)\qquad
M(t)\in G;\qquad\quad \gamma(x,0)=g(x).
\label{DefMon}\ee
$X[g]$ is obtained 
as the velocity to the curve at $t=0$,
encoded by the ${\cal G}$-valued, smooth function 
\be
\xi(x):={d\over dt} g^{-1}(x)\gamma(x,t)\Big\vert_{t=0}\,.
\label{Xidef}
\ee 
The monodromy properties of $\xi (x)$ can be derived by taking 
the derivative of the first equation in (\ref{DefMon}):
$ 
\xi^\prime(x+2\pi)=M^{-1}\xi^\prime(x)M, 
$  
and this can be solved in terms of a ${\cal  G}$-valued, 
smooth, $2\pi$-periodic 
function, $X_J\in\widetilde{{\cal  G}}$, and a constant Lie algebra element,
$\xi_0$, as follows: 
\be 
\xi(x)=\xi_0+\int_0^x dy\,
g^{-1}(y)X_J(y)g(y).
\label{XipJ}
\ee
A vector field $X$ on ${\cal  M}_{chir}$ is an assignment,
$g\mapsto X[g]$,
of a vector to every point $g\in {\cal  M}_{chir}$.
Thus it can be specified 
by the assignments $g\mapsto \xi_0[g]\in {\cal  G}$ and 
$g\mapsto X_J[g]\in\widetilde{{\cal  G}}$.
Using any curve that defines $X[g]$, 
$X$ acts on a differentiable function, $g\mapsto F[g]$, 
on ${\cal  M}_{chir}$ as  
\be 
X(F)[g]={d\over d t}F[g_t]\Big\vert_{t=0}
\qquad g_t(x)=\gamma(x,t)\,.  
\ee 
Note that the evaluation functions 
$F^x[g]:=g(x)$ and ${\cal F}^x[g]:=J(x)$ are differentiable 
with respect to any 
vector field, and their derivatives are given  by 
\be
X\big(g(x)\big)= g(x)\xi(x)
\qquad\hbox{and}\qquad
X\big(J(x)\big)=\kappa X_J(x).
\label{Xsigma}\ee 
This clarifies the meaning of $X_J$ as well. 
It is also obvious from its definition that the monodromy matrix 
yields a $G$-valued differentiable function on ${\cal  M}_{chir}$, 
$g\mapsto M= g^{-1}(x) g(x+2\pi )$,
whose derivative is characterized by the ${\cal  G}$-valued function 
\be
X(M) M^{-1} =  M\xi (x +2\pi) M^{-1} -\xi (x).
\label{XM}\ee
Having defined vector fields, one can also introduce differential 
forms as usual. 
We only remark that by (\ref{Xsigma}) evaluation 1-forms like 
$dg(x)$, $dJ(x)$ or $(g^{-1}dg)^\prime(x)$ are perfectly well-defined:
e.g. $dg(x)\big( X\big)
=X(g(x))=g(x)\xi (x)$. 

Let us now show that $\Omega_{chir}^\rho$ is weakly non-degenerate,
that is $\Omega^{\rho}_{chir}(X,Y)=0$ $\forall X$ only for $Y=0$,
 if and only if (\ref{detcond}) holds.
In order to compute 
\be
\Omega^{\rho}_{chir}(X,Y)=
\Omega_{chir}(X,Y) +  \rho(X,Y)
\label{Omega}\ee
for two arbitrary vector fields, 
we take  $X$ to be parametrized by $\xi(x)$ and further by the pair 
$\big(\xi_0,X_J(x)\big)$, while   
the analogous parametrization for $Y$ is given  by $\eta(x)$ 
and the pair $\big(\eta_0,Y_J(x)\big)$.
Then a straightforward calculation gives that 
\bea
\Omega_{chir}^\rho(X,Y)&=&\int_0^{2\pi} dx 
{\tr}\Big(X_J(x)g(x)\Big(\eta(x) + q_-(M)( M^{-1}Y(M))\Big) g^{-1}(x)\Big)
\nonumber\\
&  + & {\tr} 
\left(\xi_0\left(q_-(M) - \Ad M\circ q_+(M)\right)( M^{-1} Y(M)) \right).
\label{OmegaXY}\eea
This vanishes for every $X$, that is for arbitrary
$X_J\in \widetilde{{\cal  G}}$ and $\xi_0\in {\cal  G}$, if and only if 
\be
\left(q_-(M) - \Ad M\circ q_+(M)\right)( M^{-1} Y(M))=0,
\qquad
\eta(x) + q_-(M)( M^{-1}Y(M))=0.
\label{deger}\ee
Since the transpose with respect to the scalar product on ${\cal  G}$
satisfies 
\be  
\left(q_-(M) - \Ad M\circ q_+(M)\right)^T = 
\left(q_-(M) \circ \Ad M^{-1} - q_+(M)\right),
\ee
if (\ref{detcond}) holds then it follows from (\ref{deger}) that 
$\eta(x)$ must vanish, that is $Y=0$.
Thus we proved that (\ref{detcond}) implies the non-degeneracy of 
$\Omega_{chir}^\rho$.
The converse statement is also easy to establish, since if 
the determinant in (\ref{detcond}) vanished say at $M^0$,
then there would exist a non-zero
$A\in {\cal  G}$ such that $\left(q_-(M^0) - 
\Ad M^0\circ q_+(M^0)\right)(A)=0$.
We could hence define a tangent vector $Y^0$ at a corresponding 
point in ${\cal  M}_{chir}$
by $\eta^0(x)= - q_-(M^0)(A)$, and this vector would 
annihilate $\Omega_{chir}^\rho$.
(The definition of $Y^0$ is consistent since 
$\eta^0(x+2\pi) - (M^0)^{-1} \eta^0(x) M^0 =A$ holds.)

Now we turn to our main problem:
For a differentiable (scalar) function 
$F$ on the phase space $\check {\cal  M}_{chir}$, 
we wish to find a corresponding vector field, $Y^F$, satisfying 
\be 
X(F)=\kappa\Omega^{\rho}_{chir}(X,Y^F) 
\label{hamvect}\ee 
for all vector fields $X$. 
Notice that $Y^F$ does not necessarily exist for a given $F$. 
We say that $F$ is an element of the set of {\em admissible Hamiltonians},
denoted as ${\tt H}$, 
if the corresponding hamiltonian vector field, $Y^F$,  exists.
On account of the non-degeneracy of $\Omega_{chir}^\rho$,
if $Y^F$ exists then it is uniquely determined.

We may use the formula (\ref{OmegaXY}) for $Y:=Y^F$  
to establish the following three necessary and
sufficient conditions that $F$ must obey to guarantee that $Y^F$
exists:\hfill\break\noindent 
$\bullet$ There must exist 
a {\em smooth} ${\cal  G}$-valued function on ${\bf R}$, 
$A^F(x)$, and a
constant Lie algebra element, $a^F$, such that for any vector field $X$ 
\be
X(F)=\kappa\int_0^{2\pi}dx \tr\Big(X_J(x)A^F(x)\Big)+ 
\kappa\tr\big(\xi_0a^F\big).
\label{Egy}\ee
(This means that $F\in {\tt H}$ must have an exterior derivative 
parametrized
by the assignments $g\mapsto A^F(x)[g]$ and $g\mapsto a^F[g]$.
The restriction of $A^F(x)$ to $x\in [0,2\pi]$ and $a^F$ are
uniquely determined by (\ref{Egy}), and $A^F(x)$  is made a unique   
function on  $\bf R$ by the next requirement.) 
\hfill\break\noindent $\bullet$ The expression 
\be
\Big[A^F(x),J(x)\Big] +\kappa {d {A^F}(x)\over dx}
\label{Ketto}
\ee
must define a smooth {\em $2\pi$-periodic} function on ${\bf R}$. 
\hfill\break\noindent $\bullet$ $A^F(x)$ and $a^F$ must be related by 
 \be
a^F=g^{-1}(0)\Big[A^F(0)-A^F(2\pi)\Big]g(0).
\label{Harom}\ee 
{\em If these conditions are satisfied, then $Y^F$ is in fact given by} 
\be
g^{-1}(x) Y^F(g(x))= g^{-1}(x)A^F(x)g(x)-{1\over2}a^F+r(M)\big(a^F\big)\,,
\label{etasol}
\ee 
{\em where $r(M)$ is the linear operator on ${\cal  G}$ defined by}
\be
r(M)= {1\over 2} 
\left( q_+(M) -  q_-(M)\circ \Ad (M^{-1})\right)^{-1} 
\circ 
\left(q_+(M)+ q_-(M) \circ \Ad (M^{-1})  \right).
\label{rexplicit}\ee
Note that the matrix $r^{\alpha\beta}(M)$ of $r(M)$ is antisymmetric. 
Later we shall also use  
the operators  $r_\pm(M):= r(M) \pm {1\over 2} I$ and the corresponding 
${\cal  G}\otimes {\cal  G}$-valued functions on $\check G$:
\be
\hat r(M):= r^{\alpha\beta}(M) T_\alpha \otimes T_\beta,
\qquad
\hat r_\pm(M)=\hat r(M) \pm {1\over 2}\hat I,
\qquad
\hat I= T^\alpha \otimes T_\alpha.
\ee

The above explicit description of the hamiltonian map $F\mapsto Y^F$ 
induced by $\kappa \Omega_{chir}^\rho$ on $\check {\cal  M}_{chir}$ is one 
of our main results \cite{hosszu}.
Its proof can be sketched as follows.
First, by assuming that (\ref{hamvect}) holds we see from (\ref{OmegaXY})
for $Y= Y^F$ that at every point in the phase space $X(F)$ has the form 
(\ref{Egy}) with 
\be
g^{-1}(x)A^F(x) g(x)= \eta(x) + q_-(M)(M^{-1} Y^F(M)),
\label{P1}\ee
\be
a^F = \left(q_-(M) - \Ad M\circ q_+(M)\right)( M^{-1} Y^F(M)).
\label{P2}\ee
Since by the meaning of tangent vectors we must have 
$\eta'(x)= g^{-1}(x) Y^F_J(x) g(x)$, by taking the derivative of (\ref{P1})
we immediately get that 
\be
Y^F(J(x))= [A^F(x), J(x)] +  \kappa \partial_x {A^F}(x) 
\ee
must hold, which in particular means that the right hand side 
must define a smooth, $2\pi$-periodic function on ${\bf R}$.
As for equation (\ref{Harom}),
this is a direct consequence of (\ref{P1}) and (\ref{P2}) 
by taking into account that as a tangent vector $Y^F$ satisfies
\be
M^{-1} Y^F(M)= \eta(2\pi) - M^{-1} \eta(0) M.
\ee 
This proves that the elements of ${\tt H}$ indeed meet the
conditions (\ref{Egy}), (\ref{Ketto}), (\ref{Harom}).
Moreover, if the hamiltonian vector field exists then 
by combining (\ref{P1}) and (\ref{P2}) we obtain 
\be
g^{-1}(x) Y^F(g(x))=  g^{-1}(x)A^F(x) g(x) - q_-(M) \circ 
\left(q_-(M) - \Ad M\circ q_+(M)\right)^{-1} (a^F),
\ee
which is equivalent to (\ref{etasol}), since for the
operator $r(M)$ defined by (\ref{rexplicit})
\be
r_-(M)= - q_-(M) \circ \left(q_-(M) - \Ad M\circ q_+(M)\right)^{-1}
\ee
is an identity.
To complete the proof, one checks that if the expression in  
(\ref{Ketto}) is $2\pi$-periodic, then (\ref{etasol}) 
gives a well-defined vector field 
(since $g^{-1}(x)Y^F(g(x+2\pi))M^{-1} - g^{-1}(x) Y^F(g(x))$
is independent of $x$), which  
satisfies (\ref{hamvect}) if (\ref{Egy}) and (\ref{Harom}) hold. 
 
Now we elaborate the hamiltonian vector field for some
particular elements in ${\tt H}$. 
First note that  the matrix elements of the evaluation
functions ${\cal F}^x$ and $F^x$  fail to satisfy the first
condition, thus they are not in ${\tt H}$. However, their smeared out
versions
 \be\label{smeardef} 
{\cal F}_\mu:=\int_0^{2\pi}dx{\tr}\Big(\mu(x)J(x)\Big),\qquad 
F_\phi[g]:=\int_0^{2\pi}dx {\tr}\Big(\phi(x)g^\Lambda(x)\Big), 
\ee
(where in defining $F_\phi$ we use a 
representation\footnote{We also use the notation 
$\tr= c_\Lambda {\rm{tr}}_\Lambda$,
where $\rm{tr}_\Lambda$ is the trace over the representation 
$\Lambda$ and $c_\Lambda$ is a normalization factor that makes 
$c_\Lambda \rm{tr}(A^\Lambda B^\Lambda)$
 independent of $\Lambda$ for $A,B\in{\cal  G}$.}
$\Lambda:
G\rightarrow GL(V)$ of $G$ with $g^\Lambda =\Lambda (g)$ and 
  a smooth test function 
$\phi: {\mathbf  R} \rightarrow {\mathrm End}(V)$) can be
 shown to belong to ${\tt H}$, if 
$\mu(x)$ is a ${\cal  G}$-valued, smooth, $2\pi$-periodic test function, and
$\phi $ satisfies $\phi^{(k)}(0)=\phi^{(k)}(2\pi)=0$ for every 
integer $k\geq 0$. 
The corresponding hamiltonian vector fields obtained from (\ref{etasol}) 
satisfy 
\be
Y^{{\cal F}_\mu}\big(g(x)\big)=\mu(x)g(x),
\quad
Y^{{\cal F}_\mu}(J(x)) = [\mu(x), J(x)] + \kappa \mu'(x),
\qquad
Y^{{\cal F}_\mu}(M)=0, 
\label{YofJ}\ee
and, for $x\in [0,2\pi]$,  
\be
g^{-1}(x) Y^{F_\phi}(g(x))= 
{1\over\kappa} T^\alpha  
\int_x^{2\pi}d y \tr\left( T^\Lambda_\alpha \phi(y) g^\Lambda(y)\right)
 -{1\over 2} a^{F_\phi} + r(M)(a^{F_\phi}).
\label{Yg}\ee
Eq.~({\ref{YofJ}) shows that the ${\cal F}_\mu$ generate 
an infinitesimal action 
of the loop group on the phase space with respect to which $g(x)$ is an 
affine Kac-Moody primary field, and the current $J(x)$ 
transforms according to the co-adjoint action of the  
centrally extended loop group.
The matrix elements $M^\Lambda_{kl}$ of 
the monodromy matrix in representation
$\Lambda $ also belong to ${\tt H}$.
The action of $Y^{M^\Lambda_{kl}}$ on
$g_{ij}^\Lambda (x)$ and on $M_{ij}^\Lambda $ can be written in tensorial
form as
 \be
Y^{M^\Lambda_{kl}}\bigl( g^\Lambda_{ij}(x)\bigr)
=
{1\over \kappa} \bigl( g(x)\otimes M \, 
\hat \Theta(M)\bigr)^\Lambda_{ik, jl}\,,\label{gMPB}\ee
\be
 Y^{M^\Lambda_{kl}}(M^\Lambda_{ij}) = {1\over \kappa} \Big( (M\otimes M) 
\hat \Delta(M)  \Big)^\Lambda_{ik,jl}\,,
\label{MMPB}\ee
where 
our tensor product notation is $(K\otimes L)_{ik,jl}= K_{ij}
L_{kl}$, and 
\be
\hat \Theta(M) = \hat r_+(M)  -M_2^{-1} \hat r_-(M) M_2\,,\qquad
\hat \Delta(M) = \hat \Theta(M) - M_1^{-1} \hat \Theta(M) M_1 
\label{Delta}\ee 
with $M_1= M\otimes 1$, $M_2= 1\otimes M$. 

We now wish to rewrite the above hamiltonian vector fields 
in a symbolic notation of Poisson brackets. 
Recall that the PB of two smooth functions $F_1$ and $F_2$ 
on a {\em finite dimensional} smooth symplectic manifold is defined by 
\be
\{ F_1, F_2\} = Y^{F_2}(F_1)=-Y^{F_1}(F_2)=\Omega(Y^{F_2},Y^{F_1}),
\ee
where $Y^{F_i}$ is the hamiltonian vector field associated with $F_i$
by the symplectic form $\Omega$.
One may formally apply the same formula in the infinite dimensional
case to the `smooth enough' admissible functions.
However, it is a non-trivial problem to precisely specify the
set of functions that form a closed Poisson algebra.
Setting this question aside, 
it is clear from (\ref{YofJ}) and (\ref{MMPB}) 
that the admissible functions of $J$ and those of $M$ 
will form two closed Poisson subalgebras that centralize each other.  
Furthermore, we may  
use the perfectly well-defined expression 
\be
\{ F_\chi, F_\phi\}:= Y^{F_\phi}( F_\chi)
\ee
for the PB of two admissible Hamiltonians  
of  type $F$ in eq.~(\ref{smeardef}) to define the 
(`distribution valued') PB of
the evaluation functions $g(x)$ by the equality:
\be
 \{ F_\chi, F_\phi\}:=
\int_0^{2\pi} \int_0^{2\pi} dx dy  {\mathrm Tr}_{12}
\left( \chi(x)\otimes \phi(y) 
\{ g^\Lambda(x) \stackrel{\otimes}{,} g^\Lambda(y)\}\right),
\label{localPB}\ee
where ${\mathrm Tr}_{12}$ is the (normalized) trace over $V\otimes V$ and 
$
\{ g^\Lambda(x) \stackrel{\otimes}{,} g^\Lambda(y)\}_{ik,jl}= 
\{ g^\Lambda_{ij}(x), g^\Lambda_{kl}(y)\}$.   
With these definitions, our explicit formula of the hamiltonian vector
field $Y^{F_\phi}$ in (\ref{Yg})
is equivalent to the following quadratic `exchange algebra' type
PB for the chiral field $g(x)$:
\be 
\Big\{g^\Lambda(x)\stackrel{\otimes}{,} g^\Lambda(y)\Big\}
={1\over\kappa}\Big(g^\Lambda(x)\otimes
g^\Lambda(y)\Big)\Big(
\hat r(M) + {1\over 2} {\hat I} \,{\mathrm sign}\,(y-x)
\Big)^\Lambda, \quad 0< x, y<2\pi.
\label{xchPB}\ee  
Proceeding in the same way with
the $\{ F_\phi , M_{kl}^\Lambda\}$ PB
 as we did with the $\{ F_\chi,
F_\phi\}$ one, we conclude that the right hand side of 
(\ref{gMPB}) should be interpreted as the expression of
the $\{ g_{ij}^\Lambda (x) ,M_{kl}^\Lambda\}$ PB,
and similarly for (\ref{MMPB}). 
   
It is an open question if the admissible Hamiltonians of 
type ${\cal F}_\mu$, $F_\phi$ and 
$M^\Lambda_{kl}$ together generate a closed Poisson algebra.
Leaving this for a future study, we here only remark  that the Jacobi 
identity for three functions of type $F_\phi$
is in fact equivalent  to the following equation for $\hat r(M)$:
\be
\big[\hat r_{12}(M),\hat r_{23}(M)\big]
+\Theta_{\alpha\beta}(M)T^\alpha_1{\cal R}^\beta \hat r_{23}(M)
+ \hbox{cycl. perm.}=
-{1\over 4} \hat f,
\label{GCDYB}\ee
where $\hat f$ is defined by 
\be
\hat f:= f_{\alpha \beta}^\gamma T^\alpha \otimes T^\beta \otimes T_\gamma,
\qquad 
[T_\alpha, T_\beta]=f_{\alpha\beta}^\gamma T_\gamma,
\ee
and the cyclic permutation is over the three tensorial factors
with  
$\hat r_{23} = r^{\alpha\beta}  (1\otimes T_\alpha\otimes T_\beta)$,
$T^\alpha_1= T^\alpha \otimes 1 \otimes 1$ and so on.
We use the components of  
$\hat \Theta=\Theta_{\alpha\beta}T^\alpha\otimes T^\beta$
given by (\ref{Delta}), and the left-invariant differential
operators ${\cal R}^\beta$ that act on a function $\psi$ of $M$ by 
\be
({\cal R}^\beta \psi)(M):= {d\over d t}
 \psi(Me^{t T^\beta} )\Big\vert_{t=0}.
\label{derLR}\ee
Eq.~(\ref{GCDYB}) can be viewed as a dynamical generalization of the 
classical modified Yang-Baxter equation, to which it reduces 
if the r-matrix is a monodromy independent constant.
As a consequence of $d\Omega^{\rho}_{chir}=0$,
(\ref{GCDYB}) is satisfied for any $\hat r(M)$ given by (\ref{rexplicit}).

What are the Poisson-Lie (PL) symmetries of the chiral WZNW phase space?
To make this question more definite, 
we equip the group $G=\{ h\} $ with a PL structure 
by means of the Sklyanin bracket 
\be
 \{ h \stackrel{\otimes}{,} h\}_{\hat R} ={1\over \kappa} [h\otimes h, \hat R],
\label{Sklyanin}\ee
where $\hat R=R^{\alpha\beta}T_\alpha\otimes 
T_\beta\in {\cal  G}\wedge {\cal  G}$
 is a {\em constant} r-matrix satisfying    
\be
[\hat R_{12}, \hat R_{23}] +\hbox{cycl. perm.} = -\nu^2 \hat f
\label{nnu}\ee
for some constant $\nu$.
We then seek the conditions on $\hat r(M)$ and $\hat R$ that
guarantee the standard right 
action\footnote{Since $M\mapsto h^{-1}Mh$,  
we here have to assume that $\check G\subset G$ is invariant under 
the adjoint action of $G$, or should restrict our attention 
to the corresponding ${\cal  G}$-action.}    
of $G$ on $\check {\cal  M}_{chir}$, 
\be
\check {\cal  M}_{chir}\times G\ni (g, h)\mapsto gh 
\in \check {\cal  M}_{chir},
\label{rigid}\ee
to be a PL action. This leads to the requirement  
\be
\hat r(h^{-1}Mh)-\hat R = (h\otimes h)^{-1} 
\bigl( \hat r(M) -\hat R\bigr) (h\otimes h),
\label{PLcondition}\ee
i.e., right multiplication is a PL symmetry  iff 
the exchange r-matrix $\hat r(M)$ is such a solution of
(\ref{GCDYB}) that the difference $(\hat r(M)- \hat R)$ is equivariant. 
We can provide such solutions explicitly in association 
with {\em any} given solution of
(\ref{nnu}).
These solutions are obtained by assuming the validity of the exponential 
parametrization for $M\in \check G$:
\be
M:=e^{2\pi \Gamma}\quad\hbox{for}\quad \Gamma \in \check {\cal G}
\subset {\cal  G} \qquad\hbox{and}\qquad 
{\cal Y}:=2\pi (\ad \Gamma ).
\ee
Any analytic function of ${\cal Y}$ is equivariant, 
and it is possible to prove \cite{hosszu}
that the r-matrix corresponding to the linear operator
\be
r(M)=\frac{1}{2}\coth\frac{ {\cal Y}}{2}-\nu\coth(\nu{\cal Y})+R 
\label{rveg}\ee
solves both (\ref{PLcondition}) and (\ref{GCDYB}) (on the domain where its
power series converges).

We end this section with some remarks on the above  formula.
First, note that for $\nu=0$ (\ref{rveg}) is
understood as the limit of the corresponding complex analytic function.
Thus for $\nu=0$ and $R=0$ it yields  
$r_0=\frac{1}{2}\coth\frac{ {\cal Y}}{2}-\frac{1}{{\cal Y}}$.
If the PB (\ref{xchPB}) on $\check {\cal  M}_{chir}$ is defined by $r_0$,
then (\ref{rigid}) is a {\em classical $G$-symmetry}. 
Second, if  $\nu =\frac{1}{2}$ then $r=R$,
which is the case of the constant exchange r-matrices \cite{FG}.
Third,  it is worth stressing that for a compact Lie algebra ${\cal  G}$
constant exchange r-matrices do not exist,  
because of the negative sign on the right hand side of (\ref{GCDYB}),
but our formula (\ref{rveg}) gives explicit solutions of (\ref{GCDYB})
also in this case using a purely imaginary $\nu$ in (\ref{nnu}).
Finally, we remark  that in the $\nu={1\over 2}$ case 
the construction of the 2-form $\rho$ that corresponds 
to the r-matrix in (\ref{rveg}) is presented in \cite{FG}, while 
in general a suitable 
local 2-form can be obtained by solving (\ref{rexplicit}) for $q$. 
Further comments are contained in \cite{hosszu}.

\section{Hamiltonian structures for diagonal monodromy}
\setcounter{equation}{0}

We now describe the hamiltonian structure  
that results by restricting the symplectic form
$\Omega^{\rho}_{chir}$ to a submanifold 
${\cal  M}_{Bloch}\subset {\cal  M}_{chir}$ 
consisting of chiral WZNW fields with  diagonal 
monodromy. The corresponding exchange algebra PB turns out
to contain the classical dynamical r-matrix (\ref{Romega}).

In this section,  let ${\cal  G}$ be either a complex 
simple Lie algebra or its normal real form, 
and $G$ a corresponding Lie group. 
Choose a Cartan subalgebra ${\cal  H}\subset {\cal  G}$ that admits the   
root space decomposition
\be
{\cal  G}={\cal  H} \oplus \sum_{\alpha\in \Phi} {\cal  G}_\alpha,
\ee
and an associated basis $H_k\in {\cal  H}$, $E_\alpha \in {\cal  G}_\alpha$ 
normalized by
$\tr\left(E_\alpha E_{-\alpha}\right) = {2\over \vert \alpha\vert^2}$.
By using this basis any $A\in {\cal  G}$ can be decomposed as  
\be
A=A^0+ A^r
\quad\hbox{with}\quad 
A^0\in {\cal  H},\quad 
A^r=\sum_{\alpha\in \Phi} E_\alpha  \tr(E^\alpha A),
\quad
E^{\alpha}:= {1\over 2}{\vert \alpha \vert ^2 } E_{-\alpha} .
\label{0rdecomp}\ee
Fix an open domain ${\cal A}\subset {\cal  H}$ which has the properties that 
$\alpha(\omega)\notin i2 \pi  {\mathbf Z}$ for any root, 
$\alpha\in \Phi\subset {\cal H}^*$,
and the map ${\cal A}\ni \omega\mapsto e^{\omega} \in G$ is injective.

Then define ${\cal  M}_{Bloch}\subset {\cal  M}_{chir}$ by
\be
{\cal  M}_{Bloch}:=\{ b \in C^\infty({\mathbf  R}, G)\,\vert\,
b(x+2\pi)= b(x) e^{ \omega},
\quad 
\omega\in {\cal A} \subset {\cal  H}\,\}.
\label{Bloch}\ee
Let ${\cal  M}_{Bloch}$ be equipped  with the 2-form 
$\kappa \Omega_{Bloch}^{\rho_B}$,
where 
\be
\Omega_{Bloch}^{\rho_B}(b):=
- {1\over 2 }\int_0^{2\pi}dx\, 
 {\tr}\left(b^{-1}d b \right) \wedge
\left(b^{-1}db \right)'
-{1\over 2 }  {\tr}\left((b^{-1} d b)(0)\wedge d \omega\right)
+ \rho_B(\omega)
\label{Blochform}\ee 
with an arbitrary {\em closed} 2-form $\rho_B$ on ${\cal A}$.
Clearly, $\Omega_{Bloch}^{\rho_B}$ could be obtained from 
$\Omega_{chir}^\rho$ (\ref{Orho}) upon imposing the constraint $M=e^\omega$.
Now $\rho_B$ is parametrized as 
\be
\rho_B(\omega) = {1\over 2} q_B^{kl}(\omega) \tr(H_k d\omega) 
\wedge \tr(H_l d\omega),
\qquad
q_B^{kl}=-q_B^{lk}, 
\ee
and a corresponding linear operator $q_B(\omega)$ on ${\cal  H}$ 
is defined  by
\be
q_B(\omega) (C)= H_k q_B^{kl}(\omega) \tr (H_l C) 
\qquad \forall C\in {\cal  H}.
\ee  

To show that $\Omega_{Bloch}^{\rho_B}$ is symplectic, 
it will be convenient to parametrize $b\in {\cal  M}_{Bloch}$ as 
\be
b(x) = h(x) \exp\left(x \bar\omega \right),
\qquad \bar \omega:= {\omega\over 2\pi},
\label{Blochpar}\ee
where $\omega\in {\cal A}$ and $h\in \widetilde{G}$.
This one-to-one parametrization yields the identification 
\be
{\cal  M}_{Bloch} =\widetilde{G}\times {\cal A}=\{ (h, \omega)\}.
\ee
Correspondingly, 
a vector field $X$ on ${\cal  M}_{Bloch}$ is parametrized by
\be
X=(X_h, X_\omega)
\qquad X_h \in T_h \widetilde{G}
\qquad
X_\omega \in T_\omega {\cal A} \simeq {\cal  H}
\ee
with $h^{-1} X_h \in T_e \widetilde{G} \simeq \widetilde{{\cal  G}}$.
By regarding $\omega$ and $h$ as evaluation functions 
on ${\cal  M}_{Bloch}$, we may write 
$X_\omega=X(\omega)$ and $X_h(x) = X(h(x))$.
Equivalently, $X$ can be characterized by its action on $b(x)$,
\be
b^{-1}(x) X(b(x))=e^{-x\bar \omega } h^{-1}(x) X(h(x)) e^{x\bar \omega} 
+ x X(\bar\omega),
\label{Xb}\ee
where the function $b^{-1}(x) X(b(x))$ on ${\bf R}$ is
uniquely determined by its restriction to $[0, 2\pi]$.
Naturally, the derivative $X(F)$ of a function $F$ on ${\cal  M}_{Bloch}$ 
is defined by using that any vector is the velocity to a smooth curve. 
That is, if the value of the vector field $X$ at $b\in {\cal  M}_{Bloch}$ 
coincides with the velocity to the curve  $\gamma (x,t)$ at $t=0$, 
$\gamma (x,0)=b(x)$, then for a 
differentiable  function $F$ we have 
$X(F)[b]=\frac{\rm d}{{\rm d}t}F[\gamma (x,t)]\vert_{t=0}$.  

Arguing similarly to section 3, it can be shown 
that $\Omega_{Bloch}^{\rho_B}$ is weakly non-degenerate on 
${\cal  M}_{Bloch}$ for any $\rho_B$.
The {\em admissible} Hamiltonians that possess  hamiltonian
vector fields now turn out to be those functions $F$ on ${\cal  M}_{Bloch}$ 
whose derivative with respect to any vector field $X$ 
exists and has the form 
\be
X(F) = \langle dF,X \rangle=\tr( d_\omega F X_\omega)+
\int_0^{2\pi} dx\, \tr\left( (h^{-1} d_h F) (h^{-1} X_h)\right)
\label{A1}\ee
where 
\be
dF=(d_h F, d_\omega F)
\quad
\hbox{with}\quad 
d_h F\in T_h^* \widetilde{G},
\quad
d_\omega F\in T^*_\omega {\cal A}
\label{A2}\ee
is the exterior derivative of $F$.
We here identify $T^*_\omega {\cal A}$ with ${\cal  H}$ by means of 
the scalar product $\tr$ and also identify $T_e^* \widetilde G$ 
with $\widetilde{{\cal  G}}$ by the scalar product 
$\int_0^{2\pi} \tr(\cdot,\cdot)$,
whereby we have $h^{-1} d_h F\in T_e^*\widetilde{G}=\widetilde{{\cal  G}}$.
It is clear that the local evaluation functions 
${\cal  M}_{Bloch}\ni b\mapsto b^\Lambda_{kl}(x)$
(with the matrix elements $b^\Lambda_{kl}$ taken in some 
representation $\Lambda$ of $G$)
are differentiable but not admissible, while e.g.~the Fourier 
coefficients of the components of $J=\kappa b' b^{-1}$
as well as the components of $\omega$ yield admissible Hamiltonians.

Next we prove that
$\kappa\Omega_{Bloch}^{\rho_B}$ indeed permits
to associate a unique hamiltonian vector field, $Y^F$,  
with any  Hamiltonian, $F$, subject to (\ref{A1}), (\ref{A2}). 
By definition, $Y^F$ must satisfy 
\be
\langle dF, X\rangle =X(F)=\kappa \Omega^{\rho_B}_{Bloch}(X,Y^F)
\label{YFdef}\ee
for any vector field $X$.
To determine $Y^F$,
we  first point out that in terms of $(h,\omega)$
\bea
\Omega_{Bloch}^{\rho_B} (h,\omega)&=&
-{1\over 2} \int_0^{2\pi}dx\, \tr\Bigl(
(h^{-1} dh)\wedge (h^{-1} dh)' \nonumber\\
 &+&  2\bar \omega (h^{-1} dh)\wedge (h^{-1} dh)
-2d\bar \omega \wedge h^{-1} dh \Bigr) + \rho_B(\omega).
\eea
By equating the coefficients of $h^{-1} X(h)$ and $X(\omega)$ 
on the two sides
of (\ref{YFdef}), we obtain the  following equations for $Y^F$: 
\bea
2\pi \left(h^{-1} Y^F(h)\right)' + [h^{-1} Y^F(h), \omega] 
+ Y^F(\omega)&=&
-{2\pi\over \kappa} h^{-1} d_h F
\label{G1}\\
2\pi q_B(\omega) (Y^F(\omega)) + \int_0^{2\pi}dx\, 
(h^{-1} Y^F(h))^0 &=& {2\pi\over \kappa}  d_\omega F.
\label{G2}\eea
Given $d_h F$ and $d_\omega F$, we will determine $b^{-1} Y^F(b)$,
which is equivalent to finding  $h^{-1} Y^F(h)$ and  $Y^F(\omega)$.

On account of (\ref{Xb}), 
(\ref{G1}) is in fact equivalent to 
\be
\left( b^{-1} Y^F(b)\right)'(x)= -{1\over  \kappa} 
e^{-\bar\omega x} (h^{-1} d_h F)(x)
e^{\bar\omega x},
\label{G4}\ee
whose solution   is 
\be
b^{-1}(x) Y^F(b(x))= b^{-1}(0)Y^F(b(0)) 
- {1\over \kappa} \int_0^x dy\, e^{-\bar\omega y} (h^{-1} d_h F)(y) 
e^{\bar\omega y}.
\label{G5}\ee
Hence the only  non-trivial problem  is to determine the initial value
\be
Q_F:= b^{-1}(0) Y^F(b(0)) = h^{-1}(0) Y^F(h(0)).
\label{G6}\ee
To this end, note from (\ref{Xb}) that
\be
Y^F(\omega)=
e^{\omega} b^{-1}(2\pi) Y^F(b(2\pi)) e^{- \omega} -b^{-1}(0) Y^F(b(0)).
\label{G8}\ee
By using (\ref{G5}), 
the Cartan part of (\ref{G8})  requires that
\be
Y^F(\omega)= -{1\over \kappa} \int_0^{2\pi} dx\, (h^{-1} d_h F)^0(x),
\label{G9}\ee
while the root part of (\ref{G8}) gives
\be
e^{-\omega} Q_F^r e^{ \omega} -Q_F^r= -{1\over \kappa} 
\int_0^{2\pi} dx\, e^{-\bar \omega x} (h^{-1} d_h F)^r(x) e^{\bar \omega x},
\label{G10}\ee
where $Q_F=Q_F^0+Q_F^r$ 
according to (\ref{0rdecomp}).
Then  (\ref{G10}) completely determines $Q_F^r$ as
\be
Q_F^r={1\over \kappa}\sum_{\alpha\in \Phi} 
{E_\alpha\over 1-e^{-\alpha(\omega)}}
\int_0^{2\pi} dx\, e^{-\alpha(\bar\omega)x} \tr
\left( (h^{-1} d_h F)(x) E^\alpha\right).
\label{G11}\ee
As for the remaining unknown, $Q_F^0$, (\ref{G2}) with (\ref{Xb}) 
and (\ref{G8}) 
leads to the result:
\be
2\pi \kappa Q_F^0=2\pi d_\omega F + 
 (2\pi q_B(\omega)- 
 \pi I )\left(\int_0^{2\pi} dx\,(h^{-1} d_h F)^0(x) \right)
+\int_0^{2\pi} dx\, \int_0^x dy\,
(h^{-1} d_h F)^0(y).
\label{G12}\ee  

In conclusion, we have found that 
the hamiltonian vector field $b^{-1} Y^F(b)$ is uniquely 
determined and is explicitly given by (\ref{G5}) with $b^{-1}(0) 
Y^F(b(0))=Q_F$ 
in (\ref{G11}), (\ref{G12}).
In the derivation of $Y^F$ we have crucially used that $\omega$ is restricted
to the domain ${\cal A}\subset {\cal  H}$.
At the excluded points of ${\cal  H}$ some denominators in (\ref{G11}) 
may vanish, whereby $\Omega_{Bloch}^{\rho_B}$ becomes singular.

The Poisson bracket of two `smooth enough' admissible Hamiltonians
$F_1$ and $F_2$ on ${\cal  M}_{Bloch}$ is determined by the formula 
$\{ F_1, F_2\}=\kappa\Omega_{Bloch}^{\rho_B}(Y^{F_2}, Y^{F_1})$ 
and now we extract
a `classical exchange algebra' from this formula.
Analogously to the previous section, for this we consider functions 
of the form 
\be
F_{\phi}(h, \omega)= \int_0^{2\pi} dx\, \tr( \phi(x) b^\Lambda(x) ),
\ee
where $b^\Lambda(x)$ is taken in a  
representation
$\Lambda$ of $G$ 
and $\phi(x)$ is a smooth, matrix valued,
smearing-function in that representation.
It is easy to check that $F_\phi$ is admissible if  
\be
\phi^{(k)}(0)=\phi^{(k)}(2\pi)= 0 \quad \forall k=0,1,2\ldots\,,
\label{G13}\ee
and the exterior derivative of $F_\phi$ at $(h,\omega)$ is given by  
\be
(d_\omega F_\phi)(h,\omega)={1\over 2\pi} 
\sum_k H^k \tr\left( H^\Lambda_k \int_0^{2\pi} dx\, 
( x \phi(x) b^\Lambda(x) )\right),
\label{G14}\ee
\be
\left( (h^{-1} d_h F_\phi)(h,\omega)\right)(x)=
\sum_a T^a \tr\left( 
\phi(x) h^\Lambda(x)T^\Lambda_a e^{x \bar\omega^\Lambda} \right)
\quad\hbox{for}\quad
x\in [0, 2\pi].
\label{G15}\ee
We here denote by $H_k$, $H^k$ and $T_a$, $T^a$ dual bases of ${\cal  H}$ 
and ${\cal  G}$, respectively.
The last formula extends to a smooth $2\pi$-periodic function 
on the real line
precisely if (\ref{G13}) is satisfied.      
The hamiltonian vector field
$Y^{F_\phi}$ is then found to be 
\be
\left( b^{-1} Y^{F_\phi}(b)\right)(x)= Q_{F_\phi} -
{1\over \kappa} \sum_a T^a \int_0^x dy\, 
\tr(\phi(y) b^\Lambda(y) T^\Lambda_a),
\quad\hbox{for}\quad
x\in [0, 2\pi],
\label{G16}\ee
where $Q_{F_\phi}$ is determined as described above.
By combining the preceding formulae,  one can verify that 
\be
\{ F_{\chi}, F_{\phi}\}=
\kappa\Omega_{Bloch}^{\rho_B}(Y^{F_{\phi}},Y^{F_{\chi}})=
\int_0^{2\pi} \int_0^{2\pi} dx dy  \tr_{12}\left( \chi(x)\otimes \phi(y) 
\{ b^\Lambda(x) \stackrel{\otimes}{,} b^\Lambda(y)\}\right)
\label{locB}\ee
holds for any $\phi$, $\chi$ subject 
to (\ref{G13}) {\em provided that one has} 
\be 
\Big\{b^\Lambda(x)\stackrel{\otimes}{,} b^\Lambda(y)\Big\}
={1\over\kappa}\Big(b^\Lambda(x)\otimes
b^\Lambda(y)\Big)\Big(\hat {\cal R}(\omega) 
+{1\over 2}{\hat I} \,{\mathrm sign}\,(y-x)\Big)^\Lambda, 
\quad 0< x,y<2\pi
\label{xcharep}\ee 
{\em with the dynamical r-matrix} 
\be
\hat {\cal R}(\omega)= {1\over 4}\sum_{\alpha\in \Phi} 
{\vert \alpha\vert^2}
\coth({1\over 2}\alpha(\omega)) 
E_\alpha\otimes E_{-\alpha} + \sum_{kl} q_B^{kl}(\omega) H_k \otimes H_l.
\label{Romega}\ee
The local formula (\ref{xcharep})
completely encodes the Poisson brackets on ${\cal  M}_{Bloch}$ 
since $Y^{F_{\phi}}$ can be recovered 
if the right hand side of  (\ref{locB}) is given.

The hamiltonian vector 
fields associated with $\omega_k := \tr(\omega H_k)$ and with the functions 
${\cal F}_\mu$ of $J$ (see (\ref{smeardef})) can be checked to be
\be
Y^{\omega_k} (b(x))={1\over \kappa} b(x) H_k,
\qquad
Y^{{\cal F}_\mu}\big(b(x)\big)=\mu(x)b(x).
\ee
Thus $J$ generates an action of the affine Kac-Moody algebra 
on ${\cal  M}_{Bloch}$ centralized by the action of ${\cal  H}$ 
generated by $\omega$.

The dynamical r-matrix (\ref{Romega}) is antisymmetric, and 
is neutral in the sense that
\be
[ H_k \otimes 1 + 1\otimes H_k, \hat {\cal R}(\omega) ]=0,
\label{neutral}\ee
which ensures the validity of the Jacobi identity for the 
three functions $F_\phi$, $F_\chi$, $\omega_k$.
Moreover, it satisfies the equation  
\be
[\hat {\cal R}_{12}(\omega), \hat {\cal R}_{23}(\omega)] +
 \sum_k H_1^k  {\partial \over \partial \omega^k} 
\hat {\cal R}_{23}(\omega) + \hbox{cycl.~perm.}  =-{1\over 4} \hat f 
\label{CDYB}\ee
that ensures the Jacobi identity for three functions of type $F_\phi$.
This  dynamical generalization of the modified classical Yang-Baxter 
equation arises in other 
contexts as well \cite{Gervais,Feld,Avan} and has been much studied recently 
\cite{EV,Liu,Lu}.
The exchange r-matrix (\ref{Romega}) of the chiral
WZNW Bloch waves was first obtained in \cite{BDF}, where
it was also shown that it satisfies (\ref{CDYB}).

\section{Poisson-Lie groupoids from chiral WZNW}
\setcounter{equation}{0}

It is well-known \cite{Dri} that one can associate 
a Poisson-Lie (PL) group with any antisymmetric solution of 
the modified classical Yang-Baxter equation.
Remarkably, the dynamical generalizations of this 
equation that arise in the WZNW model permit analogous 
interpretations in terms of PL groupoids.
In particular, this means that one can 
associate a PL groupoid with any chiral extension of the WZNW phase space.
Below we briefly describe these groupoids, which are finite 
dimensional Poisson manifolds  
that encode the (non Kac-Moody aspects of the) infinite dimensional 
chiral WZNW PBs.
 
Roughly speaking, a groupoid is a set, say $P$,  
endowed with a `partial multiplication'
that behaves similarly to a group multiplication in 
the cases when it can be performed (see e.g. \cite{Gpoid}).
In the cases of our interest   
$P= S \times G \times S = \{ (M^F, g, M^I) \}$, 
where $G$ is a group and $S$ is some set. 
The partial multiplication is defined for those triples 
$(M^F, g, M^I)$ and $(\bar M^F, \bar g, \bar M^I)$ for which 
$M^I=\bar M^F$,
and the product is 
\be
(M^F, g, M^I) (\bar M^F, \bar g, \bar M^I):= (M^F, g\bar g, \bar M^I)
\quad\hbox{for}\quad M^I=\bar M^F.
\label{Pmult}\ee 
Thus the graph of the partial multiplication
is the subset of 
\be
P\times P\times P =
\{ (M^F, g, M^I)\} \times \{ (\bar M^F, \bar g, \bar M^I)\}
\times \{ (\hat M^F, \hat g, \hat M^I)\} 
\ee
defined by the constraints
\be
M^I=\bar M^F,
\quad
\hat M^F= M^F,
\quad
\hat M^I=\bar M^I,
\quad
\hat g= g\bar g,
\label{graph}\ee
where the hatted triple encodes the components of the product. 
A PL groupoid \cite{Wei} is a (Lie) groupoid $P$, 
which is also a Poisson manifold 
in such a way that the graph of the partial
multiplication is a {\em coisotropic} submanifold of  
$P\times P\times P^-$, where $P^-$ denotes the  manifold
$P$ endowed with the opposite of the PB on  $P$.
To put it differently,  {\em the constraints that define 
the graph are first class}. 
(For $P=S\times G\times S$, this definition 
reduces to that of a PL group if $S$ consists of a single point.)

Let us first recall the definition of the PL groupoids 
that are related to the Poisson structures on ${\cal  M}_{Bloch}$.
In a more general context, these groupoids have been introduced 
in \cite{EV}. 
They  are of the form above, where  $S$ is a domain in 
the dual of a Cartan subalgebra ${\cal  H}$ of a simple Lie group $G$.
We now identify ${\cal  H}^*$ with ${\cal  H}$ and take the domain 
to be ${\cal A}\subset {\cal  H}$ considered in section 4.
For notational convenience,   
we further identify $S$ with $\exp({\cal A})\subset \exp({\cal  H})$, 
and denote 
the components of the corresponding triples as 
$M^I= \exp(\omega^I)$ and $M^F=\exp(\omega^F)$ for  
$\omega^F, \omega^I\in {\cal A}$.
Using the standard tensorial notation and a basis $H_i$ of ${\cal  H}$, 
we then define a Poisson structure on 
$P_{Bloch}:= {\cal A} \times G \times {\cal A} = \{ (\omega^F, g, \omega^I)\}$ 
as follows:
\bea
&&
\kappa \{ g_1, g_2\}_{P_{Bloch}} = g_1 g_2 \hat {\cal R} (\omega^I) - 
\hat {\cal R}(\omega^F) g_1 g_2 
\nonumber\\
&& \kappa \{ g, \omega^I_i\}_{P_{Bloch}} = g H_i \nonumber\\
&& \kappa \{ g, \omega^F_i \}_{P_{Bloch}} = H_i g
\nonumber\\
&& \{ \omega^I_i, \omega^I_j\}_{P_{Bloch}} =
 \{ \omega^F_i, \omega^F_j\}_{P_{Bloch}} =
\{ \omega^I_i, \omega^F_j\}_{P_{Bloch}} =0,
\label{BPPB}
\eea
where $\kappa$ is a constant included
for comparison purposes and 
$\omega^{I(F)}_i=\tr(H_i\omega^{I(F)})$. 
Equations (\ref{neutral}) and
(\ref{CDYB}) for $\hat {\cal R}(\omega)$  
are sufficient for the Jacobi identities of this PB
to be satisfied \cite{EV}. 
One can also check that the graph of the
partial multiplication is coisotropic.

In some sense, the PBs (\ref{BPPB}) on $P_{Bloch}$  correspond to   
the PBs on the chiral WZNW phase space ${\cal  M}_{Bloch}$. 
Motivated by this, we now define a PL groupoid which is related to 
an arbitrary chiral extension of the WZNW phase space with
generic monodromy.
In this case, using an open domain $\check G\subset G$, 
we  take $P$ to be
\be
P= \check G \times G \times \check G,
\ee
and postulate on it a PB  $\{\ ,\ \}_P$ as follows: 
\bea
&&
\kappa \{ g_1, g_2\}_P = g_1 g_2 \hat r(M^I) - 
\hat r(M^F) g_1 g_2 
\nonumber\\
&& \kappa \{ g_1, M^I_2\}_P = g_1 M_2^I \hat \Theta(M^I)
\nonumber\\
&& \kappa \{ g_1, M_2^F\}_P = M_2^F \hat\Theta(M^F) g_1
\nonumber\\
&&\kappa \{ M^I_1, M^I_2\}_P = M^I_1 M^I_2 \hat\Delta(M^I)
\nonumber\\
&&\kappa \{ M^F_1, M^F_2\}_P = - M^F_1 M^F_2 \hat\Delta(M^F)
\nonumber\\
&&\kappa \{ M^I_1, M^F_2\}_P =0.
\label{PPB}
\eea
It is easy to verify that a PB given by
the ansatz (\ref{PPB}) 
always yields a PL groupoid, since the constraints
in (\ref{graph}) are first class for any choice
of the ${\cal  G}\otimes {\cal  G}$ valued `structure functions' 
$\hat r$, $\hat \Theta$, $\hat \Delta$
on $\check G$.
Of course, the structure functions must satisfy a system
of equations for the ansatz (\ref{PPB}) to define a PB.
These equations are spelled out in \cite{hosszu}.
The important point is that, in fact, a {\em sufficient condition} for the
Jacobi identity is obtained by assuming that 
$\hat \Theta(M)$ and $\hat \Delta(M)$ are given by (\ref{Delta})   
in terms of an antisymmetric  solution $\hat r(M)$ of (\ref{GCDYB}).

We have extracted a PL groupoid from  
any chiral extension of the WZNW phase space with generic monodromy  
by taking the triple $\hat r$, $\hat \Theta$, $\hat \Delta$
that arises in the WZNW model to be the structure functions of $\{\ ,\ \}_P$.
It should be noticed that if  
$\hat r$  is non-dynamical, then the PL groupoid $P$  
carries the same information 
as the group $G$ endowed with the corresponding Sklyanin bracket.
Among our PL groupoids there are also 
those special cases for which $(\hat r- \hat R)$ satisfies
the equivariance condition  
(\ref{PLcondition}) in relation with an arbitrary constant r-matrix 
$\hat R$ subject to (\ref{nnu}).
In these cases, it is possible to define commuting left and right
PL actions of the group $G$ (endowed with the PB (\ref{Sklyanin})) 
on $P$, reflecting the corresponding PL symmetry (\ref{rigid}) 
on the chiral WZNW phase space.

We believe that it would be interesting to study the above introduced PL
groupoids further,
for example to understand their quantization and relate them 
to the quantized (chiral) WZNW conformal field theory.

\bigskip
\bigskip

\noindent
{\bf Acknowledgements.}
L.F. wishes to thank J. Harnad for support and hospitality in Montreal.
This work was supported in part by the Hungarian 
National Science Fund (OTKA) under T019917, T030099, T025120 
and by the Ministry of Education under FKFP 0178/1999, FKFP 0596/1999.

\end{document}